\documentclass[conference]{IEEEtran}

\usepackage[pdftex]{graphicx}
\graphicspath{{../pdf/}}
\DeclareGraphicsExtensions{.pdf}

\usepackage[cmex10]{amsmath}
\begin{document}

\title{Joint Successive Cancellation Decoding of Polar Codes over Intersymbol Interference Channels}

\author{\IEEEauthorblockN{Runxin Wang, Rongke Liu and Yi Hou}
\IEEEauthorblockA{School of Electronic and Information
Engineering\\
Beihang University\\
Beijing, China \\
Email: wangrunxin@ee.buaa.edu.cn, rongke{\_}liu@buaa.edu.cn, mokyy@ee.buaa.edu.cn}}

\maketitle

\begin{abstract}
Polar codes are a class of capacity-achieving codes for the binary-input discrete memoryless channels (B-DMCs). However, when applied in channels with intersymbol interference (ISI), the codes may perform poorly with BCJR equalization and conventional decoding methods. To deal with the ISI problem, in this paper a new joint successive cancellation (SC) decoding algorithm is proposed for polar codes in ISI channels, which combines the equalization and conventional decoding. The initialization information of the decoding method is the likelihood functions of ISI codeword symbols rather than the codeword symbols. The decoding adopts recursion formulas like conventional SC decoding and is without iterations. This is in contrast to the conventional iterative algorithm which performs iterations between the equalizer and decoder. In addition, the proposed SC trellis decoding can be easily extended to list decoding which can further improve the performance. Simulation shows that the proposed scheme significantly outperforms the conventional decoding schemes in ISI channels. 
\end{abstract}


%
\IEEEpeerreviewmaketitle

\section{Introduction}

\IEEEPARstart{P}{olar} codes were proposed by Ar{\i}kan as a class of capacity-achieving codes for the binary-input discrete memoryless channels (B-DMCs) \cite{polar}.
The codes perform well in many situations, such as source coding \cite{polar_source}, physical-layer security \cite{secrecy}, multiple access channel \cite{MAC}, flash memory \cite{polar_wom}, and cooperative relaying \cite{CR}.
However, there is often intersymbol interference (ISI), which leads to memory channels, in wireless communications, storage and other communication systems.
The memory property of the ISI channels increase the error propagation phenomenon occurring in the successive cancellation (SC) decoding of polar codes.

To deal with the ISI problem, interleaving and turbo equalization can be performed \cite{TE_overview}. 
However, the iteration decoding in the turbo equalization structures may obstruct the selection of the optimal frozen set of polar codes, which limits the performance.

In this paper, we propose a joint decoding method, called SC trellis decoding, to decode the polar codes for ISI channels. In contrast to the conventional iterative algorithm which performs iterations between the equalizer and decoder, the decoder combines the equalization and conventional decoding and is performed using recursion formulas like conventional SC decoding without iterations.
Only the conventional bit-reversal interleaving is exploited and the selection of the frozen set can be obtained by Monte Carlo approach \cite{polar}.
The proposed scheme significantly outperforms the conventional decoding schemes in ISI channels.
In addition, the proposed SC trellis decoding can be easily extended to list decoding, which can further improve the performance.
The decision functions of the method have the same form as the conventional SC decoding. For information vector ${u_0^{N-1}} = \left( {u_0,u_2,\ldots,u_{N-1}} \right), 0\le i \le N-1$ and received vector ${r_0^{N-1}}$, we compare the probability $P\left( {{{r_0^{N-1}}},\hat u_0^{i - 1}|u_i=0} \right)$ and $P\left( {{{r_0^{N-1}}},\hat u_0^{i - 1}|u_i=1} \right)$.
Compared with the conventional SC decoding, the SC trellis decoding introduce the trellis state variables in the recursive calculations.

The rest of this paper is organized as follow.
Section II briefly presents some preliminaries of polar codes.
The proposed SC trellis decoding algorithm is introduced in Section III.
Simulation results are discussed in Section IV.

\section{Preliminaries}

\subsection{Polar Codes}

The generator matrices of polar codes with lengths $N = 2^n,  n = 1,2,\ldots ,$ can be written as $\mathbf{G}_N = \mathbf{B}_N {\mathbf{G}_2}^{\otimes n}$, where ${\mathbf{G}_2} = \begin{pmatrix} 1 & 0 \\ 1 & 1 \end{pmatrix} $, ${^{\otimes n}}$ is the Kronecker power and $\mathbf{B}_N$ is the bit-reversal interleaving \cite{polar}.
Due to the polarization property of polar codes, the channel will be polarized into a set of good noiseless and poor noisy sub-channels, as the block length of codewords approaches infinite.
The good noiseless sub-channels transmit the information bits while the poor noisy sub-channels transmit the frozen bits, which are known to the receiver. 
The set of the index of the frozen bits is denoted by $\mathcal{F}$, which can be chosen by Monte Carlo approach or density evolution \cite{DE2} in B-DMCs. In this paper, the Monte Carlo approach is exploited for ISI channels.

Let $ u_0^{N-1} = \left( {u_0,u_1,\ldots,u_{N-1}} \right) $ denote the information vector, where $N$ is the code length, $u_i \in \mathcal{U} = \left\{ {0,1} \right\}$. Let $ c_0^{N-1} = \left( {c_0,c_1,\ldots,c_{N-1}} \right) = u_0^{N-1} \mathbf{G}_N $ denote the codeword vector. $ r_0^{N-1} = \left( {r_0,r_1,\ldots,r_{N-1}} \right)$ denotes the received vector and $ \hat{u}_i $ represents the decoded bit of $u_i$.
The likelihood function of the $i$th information bit $u_i$ is
\begin{equation}
\label{Eq: wi}
P\left( {r_0^{N-1},\hat u_0^{i - 1}|{u_i}} \right) = \sum\limits_{ {u_{i + 1}^{N-1}} \in {\mathcal{U}^{N - i - 1}}} {\frac{1}{{{2^{N - 1}}}}{P}\left( {r_0^{N-1}|u_0^{N-1}} \right)}\text{.}
\end{equation}

\subsection{SC Decoding}

As shown in \cite{polar}, the codes can be decoded by SC decoding.
The likelihood functions of $u_i$ are
$ {P\left( {{{r_0^{N-1}}},{{\hat{{u}}}_0^{i-1}}|{u_i} = 0} \right)} $ and $ {P\left( {{{r_0^{N-1}}},{{\hat{{u}}}_0^{i-1}}|{u_i} = 1} \right)} $.
Then $u_i, i \notin \mathcal{F}$ can be decoded as
\begin{equation}
\label{Eq: u}
{\hat u_i} = \left\{ \begin{gathered}
  0,{\text{\quad if }} {P\left( {{{r_0^{N-1}}},{{\hat{{u}}}_0^{i-1}}|{u_i} = 0} \right)}   \hfill \\
  \quad\quad\quad \ge {{P\left( {{{r_0^{N-1}}},{{\hat{{u}}}_0^{i-1}}|{u_i} = 1} \right)}} \\
  1,{\text{\quad otherwise}} \hfill \\ 
\end{gathered}  \right.\text{.}
\end{equation}

\begin{figure}[!t]
\centering
\includegraphics[width=3.4in]{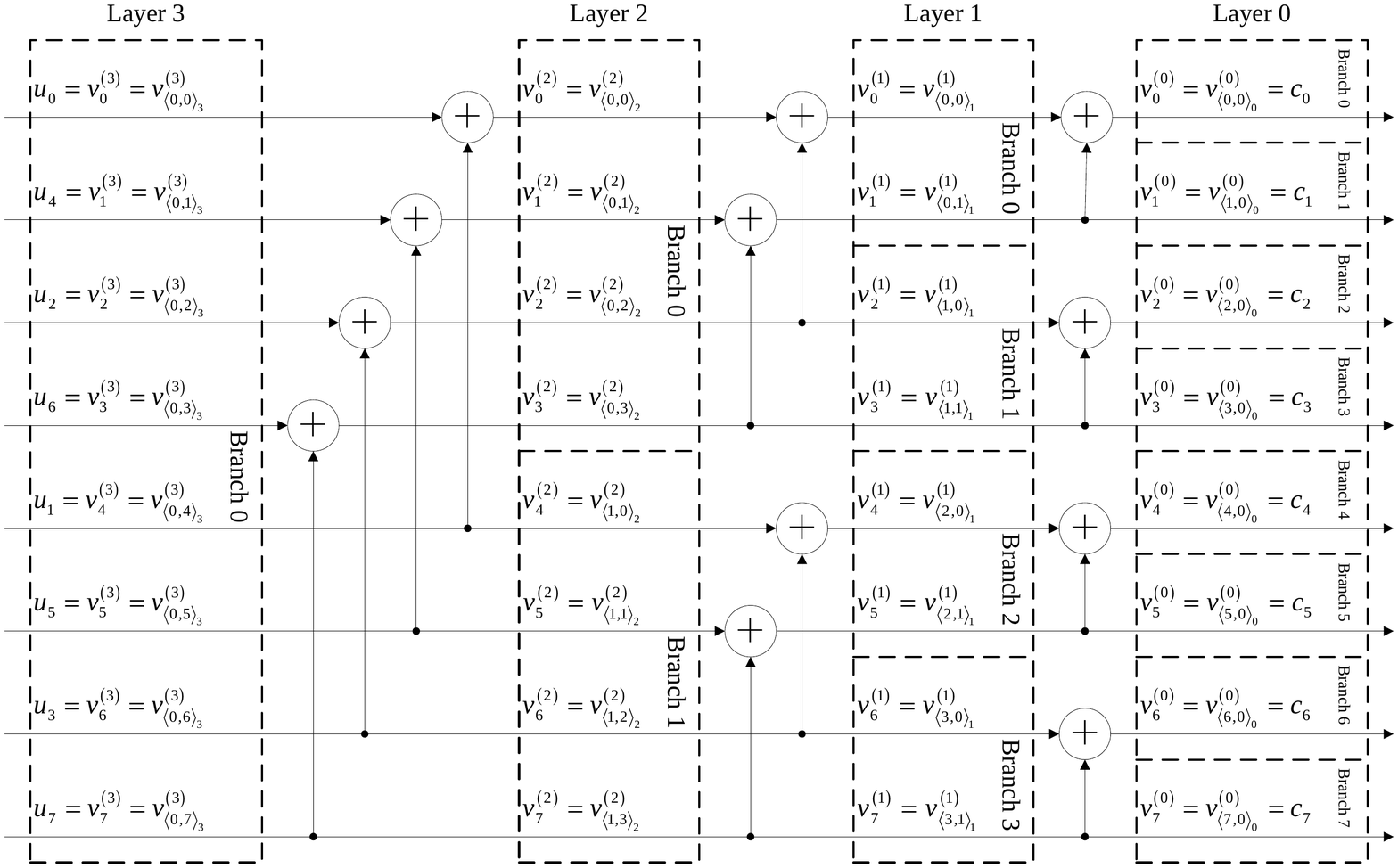}
\caption{The encoding structure of $N=8$ polar codes}
\label{Fig.: G8}
\end{figure}%

Polar codes have good encoding structures and can be decoded recursively. 
The encoder structure of $N=8$ polar codes is illustrated in Fig. \ref{Fig.: G8}.
For polar codes with code length $N=2^n$, there are $n+1$ layers in the structure.
In layer $m, 0 \le m \le n$, there are $2^{n-m}$ branches. Each branch consists of $ 2^m $ phases \cite{list}.
For convenience, we use ${\left\langle {{j_1},{j_2}} \right\rangle _m}$ to denote the summation $2^mj_1+j_2$. $v_{\left\langle {{j_1},{j_2}} \right\rangle _m}^{\left( {m} \right)}=v_{2^mj_1+j_2}^{\left( {m} \right)}$ denotes the intermediate codeword bit of $j_1$th branch and $j_2$th phase in the layer $m$. 
Note that, $v_i^{\left( {0} \right)} = c_i$ and $v_i^{\left( {n} \right)} = u_{\pi \left({i}\right)},u_i = v_{\pi^{-1} \left({i}\right)}^{\left( {n} \right)}$, where $\pi \left({\cdot}\right) = \pi^{-1} \left({\cdot}\right)$ denotes the bit-reversal interleaving.

Let $\rho _{\left( {{\left\langle {{j_1},{j_2}} \right\rangle _m},b} \right)}^{\left( m \right)}\left( {\hat u_0^{i - 1},{u_i}} \right)$ denote the summation of the likelihood functions of codeword symbols $x_{{\left\langle {{j_1},{0}} \right\rangle _m}}^{{\left\langle {{j_1},{2^m-1}} \right\rangle _m}}$ under the constraint intermediate codeword bit $v_{{\left\langle {{j_1},{j_2}} \right\rangle _m}}^{\left( {m} \right)}=b$ and fixed $u_i$, decoded $ \hat u_0^{i-1} $, i.e.,
\begin{IEEEeqnarray}{rCl}
   && \rho _{\left( {{{\left\langle {{j_1},{j_2}} \right\rangle _m}},b} \right)}^{\left( m \right)}\left( {\hat u_0^{i - 1},{u_i}} \right) \nonumber \\
&=&\sum\limits_{{\mathcal{V}}_{\left( {{{\left\langle {{j_1},{j_2}} \right\rangle _m}},b} \right)}^{\left( m \right)}\left( {\hat u_0^{i - 1},{u_i}} \right)} {P\left( {r_{{\left\langle {{j_1},{0}} \right\rangle _m}}^{{\left\langle {{j_1},{2^m-1}} \right\rangle _m}}|x_{{\left\langle {{j_1},{0}} \right\rangle _m}}^{{\left\langle {{j_1},{2^m-1}} \right\rangle _m}}} \right)} \text{,} 
\end{IEEEeqnarray}
where ${\mathcal{V}}_{\left( {{{\left\langle {{j_1},{j_2}} \right\rangle _m}},b} \right)}^{\left( m \right)}\left( {\hat u_0^{i - 1},{u_i}} \right)$ represents the condition $u_{i+1}^{N-1} \in \mathcal{U}^{N-i-1}$ with fixed $u_i$, decoded $\hat u_0^{N-1}$ and under the constraint $v_{\left\langle {{j_1},{j_2}} \right\rangle _m}^{\left( m \right)} = b$.
Note that, given an information vector $u_0^{N-1}$, all the intermediate codeword bits can be obtained uniquely, so are the codeword symbols $ x_i $. The summation $ \rho _{\left( {{\left\langle {{j_1},{j_2}} \right\rangle _m},b} \right)}^{\left( m \right)}\left( {\hat u_0^{i - 1},{u_i}} \right) $ is under the constraint $ v_{{\left\langle {{j_1},{j_2}} \right\rangle _m}}^{\left( {m} \right)}=b $, which means that given $\hat u_0^{i-1}$ and  $u_i$, all the information bits $u_{i+1}^{N-1}$ in the summation satisfy that the intermediate codeword bit $v_{{\left\langle {{j_1},{j_2}} \right\rangle _m}}^{\left( {m} \right)}=b$ according to the encoding rule.

Then the likelihood function of $u_i = v_{\pi^{-1} \left( {i} \right)}^{n}$ (\ref{Eq: wi}) can be rewritten by
\begin{IEEEeqnarray}{rCl}
   && P\left( {{r}_0^{N-1},\hat {u}_0^{i - 1}|{u_i}=b} \right) \nonumber \\
&=& \frac{1}{{{2^{N - 1}}}} \sum\limits_{{\mathcal{V}}_{\left( {{{{\pi^{-1} \left( {i} \right)}}},b} \right)}^{\left( n \right)}\left( {\hat u_0^{i - 1},{u_i}} \right)} {{P}\left( {r_0^{N-1}|x_0^{N-1}} \right)}   \nonumber \\
   &=& \frac{1}{{{2^{N - 1}}}} \rho _{\left( {\pi^{-1} \left( {i} \right),b} \right)}^{\left( n \right)}\left( { u_0^{i - 1},{u_i}} \right) \text{.} \label{Eq: P_rho}
\end{IEEEeqnarray}
Because of the good structures of polar codes, $\rho _{\left( {{\left\langle {{j_1},{j_2}} \right\rangle _m},b} \right)}^{\left( m \right)}$ can be calculated recursively like the conventional likelihood ratio calculation in \cite{polar}.

If $ j_2 <  2^{m-1}$, we have
\begin{IEEEeqnarray}{rCl}
   && \rho _{\left( {{\left\langle {{j_1},{j_2}} \right\rangle _m},b} \right)}^{\left( m \right)}\left( {\cdot} \right) \nonumber \\
   &=& \rho _{\left( {{\left\langle {{2j_1},{j_2}} \right\rangle _{m-1}},0} \right)}^{\left( {m - 1} \right)}\left(  \cdot  \right)\rho _{\left( {{\left\langle {{2j_1+1},{j_2}} \right\rangle _{m-1}},b} \right)}^{\left( {m - 1} \right)}\left(  \cdot  \right) \nonumber \\
  &\quad & +\rho _{\left( {{\left\langle {{2j_1},{j_2}} \right\rangle _{m-1}},1} \right)}^{\left( {m - 1} \right)}\left(  \cdot  \right)\rho _{\left( {{\left\langle {{2j_1+1},{j_2}} \right\rangle _{m-1}},1 - b} \right)}^{\left( {m - 1} \right)}\left(  \cdot  \right) \text{,} \label{Eq: rf_f} \\
\text{else} && \nonumber \\
   && \rho _{\left( {{j_1},{j_2},b} \right)}^{\left( m \right)}\left( {\cdot} \right) \nonumber \\
   &=& \rho _{\left( {{\left\langle {{2j_1},{j_2-2^{m-1}}} \right\rangle _{m-1}},{\hat v_{{\left\langle {{j_1},{j_2-2^{m-1}}} \right\rangle _{m}}}^{\left( {m} \right)}} \oplus b} \right)}^{\left( {m - 1} \right)}\left(  \cdot  \right)\nonumber \\
  & \quad & \quad \rho _{\left( {{\left\langle {{2j_1+1},{j_2-2^{m-1}}} \right\rangle _{m-1}},b} \right)}^{\left( {m - 1} \right)}\left(  \cdot  \right) \text{.} \label{Eq: rf_g}
\end{IEEEeqnarray}
The initialization information is the likelihood function of $x_i$, i.e., $\rho _{\left( {i,b} \right)}^{\left( {0} \right)}\left(  \cdot  \right) = P\left( {r_i|c_i=b} \right) = P \left( { {r_i|x_i = \left( {-1} \right)^{c_i}}} \right)$.

For example, let code length $N=2$, for the information bits $u_0$ and $u_1$, (\ref{Eq: rf_f}) and (\ref{Eq: rf_g}) become
\begin{IEEEeqnarray}{rCl}
   && \rho _{\left( {{\left\langle {{0},{0}} \right\rangle _{1}},b} \right)}^{\left( 1 \right)}\left( {u_0=b} \right) \nonumber \\
    &=& 2 P\left( {r_0r_1|{u_0}=b} \right) \nonumber \\
   &=& \rho _{\left( {{\left\langle {{0},{0}} \right\rangle _{0}},0} \right)}^{\left( {0} \right)}\left(  \cdot  \right)\rho _{\left( {{\left\langle {{1},{0}} \right\rangle _{0}},b} \right)}^{\left( {0} \right)}\left(  \cdot  \right) \nonumber \\
  &\quad & +\rho _{\left( {{\left\langle {{0},{0}} \right\rangle _{0}},1} \right)}^{\left( {0} \right)}\left(  \cdot  \right)\rho _{\left( {{\left\langle {{1},{0}} \right\rangle _{0}},1 - b} \right)}^{\left( {0} \right)}\left(  \cdot  \right) \nonumber \\
  &=& P\left( {r_0|{c_0}=0} \right)P\left( {r_1|{c_1}=b} \right) \nonumber \\
  &\quad & + P\left( {r_0|{c_0}=1} \right)P\left( {r_1|{c_1}=1-b} \right) \text{,} \\
\nonumber \\
   && \rho _{\left( {{\left\langle {{0},{1}} \right\rangle _{1}},b} \right)}^{\left( 1 \right)}\left( {\hat u_0,u_1 =b} \right) \nonumber \\
    &=& 2 P\left( {r_0r_1,{\hat u_0}={\hat v_{0}^{\left( {1} \right)}}|{u_1}=b} \right) \nonumber \\
   &=& \rho _{\left( {{\left\langle {{0},{0}} \right\rangle _{0}},{\hat v_{0}^{\left( {1} \right)}}\oplus b} \right)}^{\left( {0} \right)}\left(  \cdot  \right)\rho _{\left( {{\left\langle {{1},{0}} \right\rangle _{0}},b} \right)}^{\left( {0} \right)}\left(  \cdot  \right)  \nonumber \\
   &=& P\left( {r_0|{c_0}= {\hat u_0} \oplus b} \right)P\left( {r_1|{c_1}=b} \right)\text{.} 
\end{IEEEeqnarray}

\subsection{ISI Channels}

Let ${x_0^{N-1}} = \left( {x_0,x_1,\ldots,x_{N-1}} \right) \in \mathcal{X}^N = \left\{ {1,-1} \right\}^N, x_i = \left( {-1} \right)^{c_i}$ and ${r_0^{N-1}} = \left( {r_0,r_1,\ldots,r_{N-1}} \right)$ represent the codeword symbols and received vectors, respectively.
The $i$th received symbol $ r_i $ can then be written as
\begin{equation}
{r_i} = {y_i} + {n_i} = \sum\limits_{j = 0}^D {{h_j}{x_{i - j}}}  + {n_i}\text{,}
\end{equation}
where $D$ is the memory length of ISI, $n_i$ is the Gauss noise,
and $y_i \in \mathcal{Y}$ is an intermediate variable, called ISI codeword symbol in this paper.
$h_i$ is the tap coefficient of the channel.
When $D=0$, the channel becomes the binary-input additive white Gaussian noise (BI-AWGN) channel.

\begin{figure}[!t]
\centering
\includegraphics[width=3in]{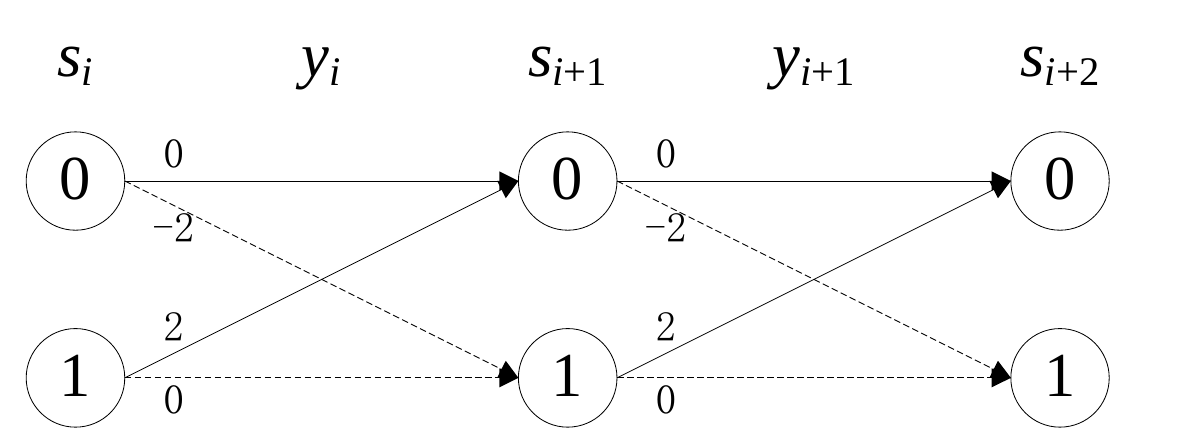}
\caption{The trellis diagram of dicode channel}
\label{Fig.: dicode}
\end{figure}

The ISI channels can also be represented by trellis diagrams. 
Let $s_i = {x_{i-D}^{i-1}} \in \mathcal{S}$ denote the trellis state.
Then in each stage there are $2^D$ states in the trellis.
For example, in dicode channel $h_0 = 1,h_1 = -1$ and there are two states in each stage. The trellis diagram of the dicode channel is showed in Fig. \ref{Fig.: dicode}.

\section{SC Trellis Decoding}

When performing soft decoding, estimations of the codeword symbols $x_i$ are required for initialization.
In B-DMCs, the likelihood function of each codeword symbol $P \left( {r_i|x_i} \right)$ is independent. On the other hand, in ISI channels, the estimations are usually obtained by equalization, and these estimations of the codeword symbols are not independent any more due to the channel memory.

To overcome the problem, in this section, we propose SC trellis decoding method for polar codes in ISI channels.
We use likelihood functions of the ISI codeword symbol $y_i$ as the initialization information and combine the processes of equalization and decoding.
This method is calculated recursively and without iteration.

\subsection{Initialization Information}

In this subsection, we will show how to use the independent likelihood functions of $y_i$ as initialization information.

If $D=0$, the ISI channel becomes BI-AWGN channel. Let $h_0 = 1$, then $r_i = y_i + n_i = h_0x_i +n_i = \left(-1\right)^{c_i} + n_i$.
The likelihood functions of the ISI codeword symbols $y_i$ are equal to the likelihood functions of the codeword symbols $x_i$, $P \left( {r_i|y_i} \right) = P \left( {r_i|x_i} \right)$, due to the bijective mapping between set $\mathcal{X}$ and $\mathcal{Y}$, i.e. $y_i = x_i = \left(-1\right)^{c_i}$.

However, if $D \ne 0$, although the bijective mapping between $\mathcal{U}$ and $\mathcal{X}$ is also satisfied ($x_i = \left(-1\right)^{c_i}$), the bijective mapping between each codeword symbol $x_i \in \mathcal{X}$ and symbol $y_i \in \mathcal{Y}$ is not satisfied any more.
For example, in the dicode channel, all the mapping between $x_i$ and $y_i$ are showed in Table \ref{Table: dicode}.
\begin{table}[!t]
\renewcommand{\arraystretch}{1.3}
\caption{the mapping between $\mathcal{X}$ and $\mathcal{Y}$ in the dicode channel, $N=2$}
\label{Table: dicode}
\centering
\begin{tabular}{|c|c|c|c|}
\hline
input bits $ {c_i^{i+1}} $ & $ {x_i^{i+1}} $ & initial state & $ {y_i^{i+1}} $ \\
\hline
$\left( {0,0} \right)$ & $\left( {1,1} \right)$ & $0$ & $\left( {0,0} \right)$ \\
\hline
$\left( {0,1} \right)$ & $\left( {1,-1} \right)$ & $0$ & $\left( {0,-2} \right)$ \\
\hline
$\left( {1,0} \right)$ & $\left( {-1,1} \right)$ & $0$ & $\left( {-2,2} \right)$ \\
\hline
$\left( {1,1} \right)$ & $\left( {-1,-1} \right)$ & $0$ & $\left( {-2,0} \right)$ \\
\hline
$\left( {0,0} \right)$ & $\left( {1,1} \right)$ & $1$ & $\left( {2,0} \right)$ \\
\hline
$\left( {0,1} \right)$ & $\left( {1,-1} \right)$ & $1$ & $\left( {2,-2} \right)$ \\
\hline
$\left( {1,0} \right)$ & $\left( {-1,1} \right)$ & $1$ & $\left( {0,2} \right)$ \\
\hline
$\left( {1,1} \right)$ & $\left( {-1,-1} \right)$ & $1$ & $\left( {0,0} \right)$ \\
\hline
\end{tabular}
\end{table}%
Note that the same input symbols $x_i$ leads different $y_i$, such as ${x_i^{i+1}} = \left( {1,1} \right)$, ${y_i^{i+1}}$ can be $ \left( {0,0} \right) $ or $ \left( {2,0} \right) $.
However, if the state $ s_i=x_{i-D}^{i-1} \in \mathcal{S} $ is fixed, we can obtain the bijective mapping between each codeword symbol $x_i \in \mathcal{X}$ and symbol $y_i \in \mathcal{Y}$ again, i.e., there is bijective mapping between $\mathcal{X} \times \mathcal{S}$ and $\mathcal{Y}$.

We consider the likelihood functions with state $s_i \in \mathcal{S}$. Then we can obtain that $P\left( {{r_{i}}|s_i,{x_{i}}} \right) = P\left( {{r_{i}}|{y_{i}}} \right)$. Because of the bijective mapping, instead of using the likelihood functions of the codeword symbol $x_i$, we can use the likelihood functions of the codeword symbol $x_i$ and state $s_i$, i.e. the likelihood functions of the ISI codeword symbols $y_i$, as the initialization information.
Fortunately, the ISI channels only have finite trellis states.
For each $i$, we can calculate the likelihood functions of $y_i$ based on all the possible values of $x_i$ and $s_i$.

\subsection{Recursive Formulas}

Note that, the likelihood functions of ISI codeword symbols $P \left( {r_i|y_i} \right)$ are independent, but unlike decoding polar codes in B-DMCs, we cannot directly use the recursive formulas (\ref{Eq: rf_f}) and (\ref{Eq: rf_g}) due to correlation between ${x_0^{N-1}}$ and ${r_0^{N-1}}$.

In ISI channels, in order to corresponding to the likelihood functions of $y_i$, the initialization information of the decoding, we consider the probability $P \left( {r_0^{N-1},{\hat u}_0^{i-1},s_0|u_i = b} \right)$ instead of $P \left( {r_0^{N-1},{\hat u}_0^{i-1}|u_i = b} \right)$. Here, $s_0$ denotes the initial state of the ISI channel. Usually, we can assume $ s_0 = 0 $ in the presence of zero padding.
Intuitively, we can also decode the codes recursively like (\ref{Eq: rf_f}) and (\ref{Eq: rf_g}).
Note that, we have 
\begin{IEEEeqnarray}{rCl}
&& P\left( {{r_0^{N-1}}|{y_0^{N-1}}} \right) \nonumber \\
&=& \prod\limits_{i=0}^{N-1} {P\left( {{r_i}|{y_i}} \right)}  \nonumber \\
&=& \prod\limits_{i=0}^{N-1} {P\left( {{r_i}|{s_i},{x_i}} \right)} = \prod\limits_{i=0}^{N-1} {P\left( {{r_i}|{s_i},{x_i},{s_{i+1}}} \right)}  \nonumber \\
&=& P\left( {r_0^{\tfrac{N}{2}-1}|{s_0},x_0^{\tfrac{N}{2}-1},s_{\tfrac{N}{2}}} \right)\nonumber \\
& \quad & \quad P\left( {r_{\tfrac{N}{2}}^{N-1}|{s_{\tfrac{N}{2} }},x_{\tfrac{N}{2}}^{N-1}, s_{N}} \right) \text{.}
\end{IEEEeqnarray}
The difference between using the likelihood functions of $y_i$ and $x_i$ is that in addition to $x_i$, a state value $s_i$ is given. Note that, if $s_i$ and $x_i$ are fixed, we can obtain the end state $s_{i+1}$ uniquely, so we rewrite ${P\left( {{r_i}|{s_i},{x_i}} \right)}$ by ${P\left( {{r_i}|{s_i},{x_i},{s_{i+1}}} \right)}$. We can divide the vector $x_i^j$ into two parts $x_i^k$ and $x_{k+1}^j$. If the end state $s_{k+1}$ of the first part is the same with the initial state of the second part, we have
\begin{IEEEeqnarray}{rCl}
\label{Eq: rf_s}
&& P\left( {r_i^j|{s_i},x_i^j,{s_{j+1}}} \right) \nonumber \\
 &=& P\left( {r_i^{k}|{s_i},x_i^{k},{s_{k+1}}} \right) \nonumber \\
& \quad  &\quad P\left( {r_{k + 1}^j|{s_{k + 1}},x_{k + 1}^j,{s_{j + 1}}} \right) \text{,}
\end{IEEEeqnarray}
where $i<k<j$. For example, in dicode channel, we have
\begin{IEEEeqnarray}{rCl}
&& P\left( {{r_i}{r_{i + 1}}|{y_i} = 0,{y_{i + 1}} =  - 2} \right) \nonumber \\
&=& P\left( {{r_i}|{s_i} = 0,{x_i} = 0,{s_{i + 1}} = 0} \right)\nonumber \\
& \quad & \quad P\left( {{r_{i + 1}}|{s_{i + 1}} = 0,{x_{i + 1}} = 1,{s_{i + 2}} = 1} \right) \text{.}
\end{IEEEeqnarray}

Let $\rho _{\left( {{\left\langle {{j_1},{j_2}} \right\rangle _m},b} \right)}^{\left( m \right)}\left( {\hat u_0^{i - 1},{u_i},s_{{\left\langle {{j_1},{0}} \right\rangle _m} },s_{{\left\langle {{j_1},{2^m}} \right\rangle _m}}} \right)$ denote the summation of the likelihood functions of ISI codeword symbols $y_{{\left\langle {{j_1},{0}} \right\rangle _m}}^{{\left\langle {{j_1},{2^m-1}} \right\rangle _m}}$ with fixed $u_i$, decoded $ \hat u_0^{i-1} $, intermediate codeword bit $v_{{\left\langle {{j_1},{j_2}} \right\rangle _m}}^{\left( {m} \right)}=b$, initial state $s_{{\left\langle {{j_1},{0}} \right\rangle _m}}$, end state $s_{ {{\left\langle {{j_1},{2^m}} \right\rangle _m}} }$, i.e.,
\begin{IEEEeqnarray}{rCl}
   && \rho _{\left( {{\left\langle {{j_1},{j_2}} \right\rangle _m},b} \right)}^{\left( m \right)}\left( {\hat u_1^{i - 1},{u_i},s_{ {\left\langle {{j_1},{0}} \right\rangle _m} },s_{{\left\langle {{j_1},{2^m}} \right\rangle _m} }} \right) \nonumber \\
   &=& \sum\limits_{{\mathcal{V}}_{\left( {{{\left\langle {{j_1},{j_2}} \right\rangle _m}},b} \right)}^{\left( m \right)}\left( {\hat u_0^{i - 1},{u_i}} \right)} {P\left( {r_{{\left\langle {{j_1},{0}} \right\rangle _m}}^{{\left\langle {{j_1},{2^m-1}} \right\rangle _m}}|{s_{{\left\langle {{j_1},{0}} \right\rangle _m}}},}\right.}\nonumber \\
  &\quad & \quad \quad \quad \quad \quad \quad \quad \quad \quad \quad  {\left.{ x_{{\left\langle {{j_1},{0}} \right\rangle _m}}^{{\left\langle {{j_1},{2^m-1}} \right\rangle _m}},{s_{{\left\langle {{j_1},{2^m}} \right\rangle _m}}}} \right)} \text{.}
  \label{Eq: rho_s} 
\end{IEEEeqnarray}
Like (\ref{Eq: P_rho}), we have
\begin{IEEEeqnarray}{rCl}
   && P\left( {{r}_0^{N-1},\hat {u}_0^{i - 1},s_0,s_N|{u_i}=b} \right) \nonumber \\
   &=& \frac{1}{{{2^{N - 1}}}} \rho _{\left( {\pi^{-1} \left( {i} \right),b} \right)}^{\left( n \right)}\left( { u_0^{i - 1},{u_i},s_0,s_N} \right) \text{.} \label{Eq: P_rho_s}
\end{IEEEeqnarray}
For the summation $\rho _{\left( {{\left\langle {{j_1},{j_2}} \right\rangle _m},b} \right)}^{\left( m \right)}\left( {\hat u_1^{i - 1},{u_i},s_{{\left\langle {{j_1},{0}} \right\rangle _m} },s_{{\left\langle {{j_1},{2^m}} \right\rangle _m}}} \right)$, we can group the terms by middle states $s_{{\left\langle {{j_1},{2^{m-1}}} \right\rangle _m}}$, i.e.,
\begin{IEEEeqnarray}{rCl}
   && \rho _{\left( {{\left\langle {{j_1},{j_2}} \right\rangle _m},b} \right)}^{\left( m \right)}\left( {\hat u_0^{i - 1},{u_i},s_{{\left\langle {{j_1},{0}} \right\rangle _m} },s_{{\left\langle {{j_1},{2^m}} \right\rangle _m}}} \right) \nonumber \\
   &=& \sum\limits_{{s_{{\left\langle {{j_1},{2^{m-1}}} \right\rangle _m}}}} \rho _{\left( {{\left\langle {{j_1},{j_2}} \right\rangle _m},b} \right)}^{\left( m \right)}\left( {\hat u_0^{i - 1},{u_i},} \right. \nonumber \\
  & \quad & \quad \quad \quad \quad \quad \quad \left.{ s_{{\left\langle {{j_1},{0}} \right\rangle _m} }, s_{{\left\langle {{j_1},{2^{m-1}}} \right\rangle _m} },s_{{\left\langle {{j_1},{2^m}} \right\rangle _m}}} \right)  \text{.}
\end{IEEEeqnarray}
Using (\ref{Eq: rf_s}), (\ref{Eq: rho_s}) and the recursive formulas (\ref{Eq: rf_f}) and (\ref{Eq: rf_g}), we can obtain the new recursive formulas.

If $ j_2 <  2^{m-1}$, we have
\begin{IEEEeqnarray}{rCl}
   && \rho _{\left( {{\left\langle {{j_1},{j_2}} \right\rangle _m},b} \right)}^{\left( m \right)}\left( {\cdot} \right) \nonumber \\
   &=& \sum\limits_{{s_{{\left\langle {{j_1},{2^{m-1}}} \right\rangle _m}}}} \left(  \rho _{\left( {{\left\langle {{2j_1},{j_2}} \right\rangle _{m-1}},0} \right)}^{\left( {m - 1} \right)}\left(  \cdot  \right)\rho _{\left( {{\left\langle {{2j_1+1},{j_2}} \right\rangle _{m-1}},b} \right)}^{\left( {m - 1} \right)}\left(  \cdot  \right) \right.\nonumber \\
  &\quad &\quad\quad\quad \left. +\rho _{\left( {{\left\langle {{2j_1},{j_2}} \right\rangle _{m-1}},1} \right)}^{\left( {m - 1} \right)}\left(  \cdot  \right)\rho _{\left( {{\left\langle {{2j_1+1},{j_2}} \right\rangle _{m-1}},1 - b} \right)}^{\left( {m - 1} \right)}\left(  \cdot  \right) \right) \text{,}\nonumber \\ \label{Eq: rf_sf} \\
\text{else} && \nonumber \\
   && \rho _{\left( {{\left\langle {{j_1},{j_2}} \right\rangle _{m}},b} \right)}^{\left( m \right)}\left( {\cdot} \right) \nonumber \\
   &=& \sum\limits_{{s_{{\left\langle {{j_1},{2^{m-1}}} \right\rangle _{m}}}}} \left( \rho _{\left( {{\left\langle {{2j_1},{j_2-2^{m-1}}} \right\rangle _{m-1}},{\hat v_{{\left\langle {{j_1},{j_2-2^{m-1}}} \right\rangle _{m}}}^{\left( {m} \right)}} \oplus b} \right)}^{\left( {m - 1} \right)}\left(  \cdot  \right) \right.\nonumber \\
   &\quad &\quad\quad\quad\quad\quad\quad\quad\quad \quad \quad    \left.  \rho _{\left( {{\left\langle {{2j_1+1},{j_2-2^{m-1}}} \right\rangle _{m-1}},b} \right)}^{\left( {m - 1} \right)}\left(  \cdot  \right) \right) \text{.} \nonumber \\
   \label{Eq: rf_sg}
\end{IEEEeqnarray}

(\ref{Eq: rf_sf}) and (\ref{Eq: rf_sg}) is the new recursive formulas instead of (\ref{Eq: rf_f}) and (\ref{Eq: rf_g}).
Introducing the middle states, the probability of the information $u_i$ can be calculated over specific states and recursively like conventional SC decoding.
The initialization information of the recursive calculations is the likelihood functions of ISI codesword symbol $y_i$, i.e., $\rho _{\left( {i,b} \right)}^{\left( {0} \right)}\left(  \cdot  \right) = P\left( {r_i|s_i,x_i,s_{i+1}} \right) = P \left( { {r_i|y_i}} \right)$.

The SC trellis decoding criterion for ISI channel is the same with (\ref{Eq: u}), i.e. we need to compare $P\left( {{{r_0^{N-1}}},{\hat {u}}_0^{i - 1}|{u_i} = b} \right)$ where $b\in \mathcal{U}=\left\{0,1\right\}$ .
So after calculate the probability $P\left( {r_0^{N-1},\hat u_0^{i - 1},s_0,s_{N}|{u_i}=b} \right)$ by (\ref{Eq: P_rho_s}) and the new recursive formulas, we need to merge the results.
We assume the initial states are 0,
so the probability can be rewritten as
$P\left( {{{r_0^{N-1}}},{\hat {u}}_0^{i - 1},{s_0} = 0|{u_i} = b} \right),b\in \mathcal{U}$, which can be calculated by
\begin{IEEEeqnarray}{rCl}
  && P\left( {{{r_0^{N-1}}},{\hat {u}}_0^{i - 1},{s_0} = 0|{u_i} = b} \right)  \nonumber \\
& = & \sum\limits_{{s_{N}}} {P\left( {{{r_0^{N-1}}},{\hat {u}}_0^{i - 1},{s_0} = 0,{s_{N}}|{u_i} = b} \right)} \text{.} 
\end{IEEEeqnarray}

\subsection{List Decoding}

SC decoding can be improved by list decoding \cite{list}. The proposed method can also be improved by list decoding.
We calculate the probability of sequences $ P\left( {r_0^{N-1}|u_0^i = b_0^i} \right) , b_i \in \mathcal{U}$, and keep the best $L$ paths. Note that, we have the following relationship.
\begin{equation}
P\left( {r_0^{N-1}|u_0^i=b_0^i} \right) = {2^{i}}P\left( {r_0^{N-1},u_0^{i - 1}=b_0^{i-1}|{u_i=b_i}} \right) \text{.}
\end{equation}
For each $0 \le i \le N-1$, we keep the best $L$ paths, who have the maximum values of the probability $P\left( {r_0^{N-1},u_0^{i - 1}|{u_i}} \right)$. For $1 \le l \le L$, there is an estimation of $u_0^{i-1}$, denoted by $\left( {b_0^{\left( {l} \right)}},{b_1^{\left( {l} \right)}},\ldots,{b_{i}^{\left( {l} \right)}} \right)$. And for $i+1$, we calculate the new $2L$ paths using the estimation $\left( {b_0^{\left( {l} \right)}},{b_1^{\left( {l} \right)}},\ldots,{b_{i}^{\left( {l} \right)}} \right)$ of the last $L$ paths and keep the best $L$ paths.

Although the SC trellis decoding requests the probability with trellis states, the decision functions $P\left( {r_0^{N-1},u_0^{i - 1},s_0|{u_i}} \right)$ are the same with the probability $P\left( {r_0^{N-1},u_0^{i - 1}|{u_i}} \right)$ if $s_0$ is fixed. It is straightforward to change SC trellis decoding to list decoding. 
We show the performance in Section \ref{Simulation}.

\section{Simulation}
\label{Simulation}
In this section, 
we compare the performance of the proposed decoding method with the turbo equalization structures of polar codes.
In the turbo equalization structures, first, BCJR equalization is performed, then it passes soft estimates to the decoder after deinterleaving. If the decoder is belief propagation (BP) decoder, it operates for $I$ iterations. The resulting soft estimates are passed back to equalizer after interleaving. This forms one turbo iteration, and is repeated $I_0$ times.
We also compare the performance with LDPC codes decoded with the turbo equalization structure using BP decoding \cite{LDPC_PR}. The LDPC codes are with column weight 3.
We choose EPR4 channel to show the performance. The channel is modelled by
\begin{IEEEeqnarray}{rCl}
{r_i} &=& {y_i} + {n_i} \nonumber \\ 
&=& x_i+x_{i-1}-x_{i-2}-x_{i-3} + {n_i} \text{.}
\end{IEEEeqnarray}

\begin{figure}[!t]
\centering
\includegraphics[width=3in]{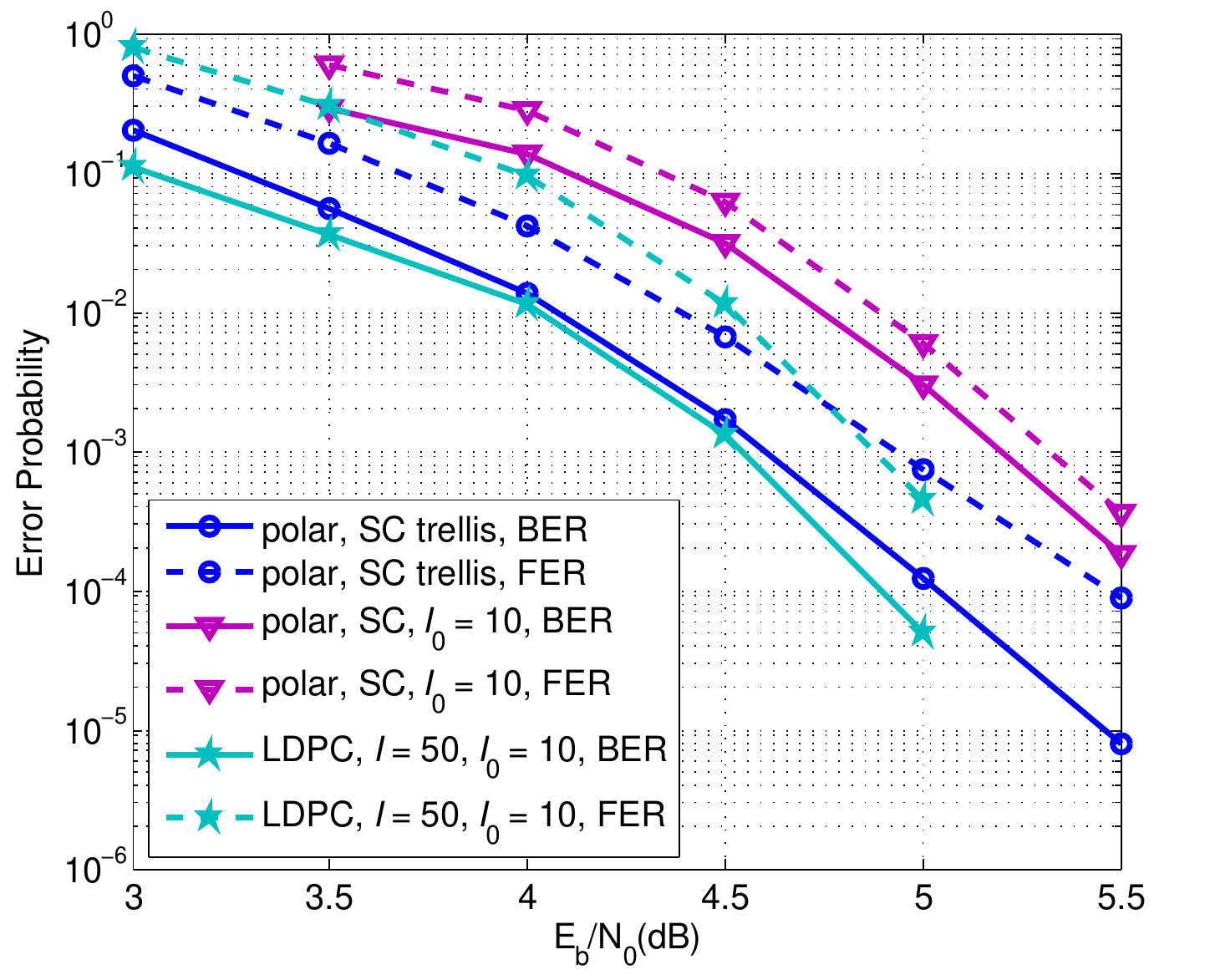}
\caption{The performance of polar codes under SC trellis decoding and SC decoding and LDPC codes under BP decoding with code length $N=1024$, code rate $R = 1/2$}
\label{Fig.: EPR4_N1024}
\end{figure}

Fig. \ref{Fig.: EPR4_N1024} shows the performance of codes with code lengths $N=1024$ and code rates $R=1/2$. The proposed SC trellis decoding method is compared with the turbo equalization structure using SC decoding.
We can find that the proposed method significantly outperforms the turbo equalization structure and can obtain 1 dB gain at BER $10^{-4}$.
Note that, the bit error rate (BER) of turbo equalization structure is close to the frame error rate(FER) due to the critical error propagation.

\begin{figure}[!t]
\centering
\includegraphics[width=3in]{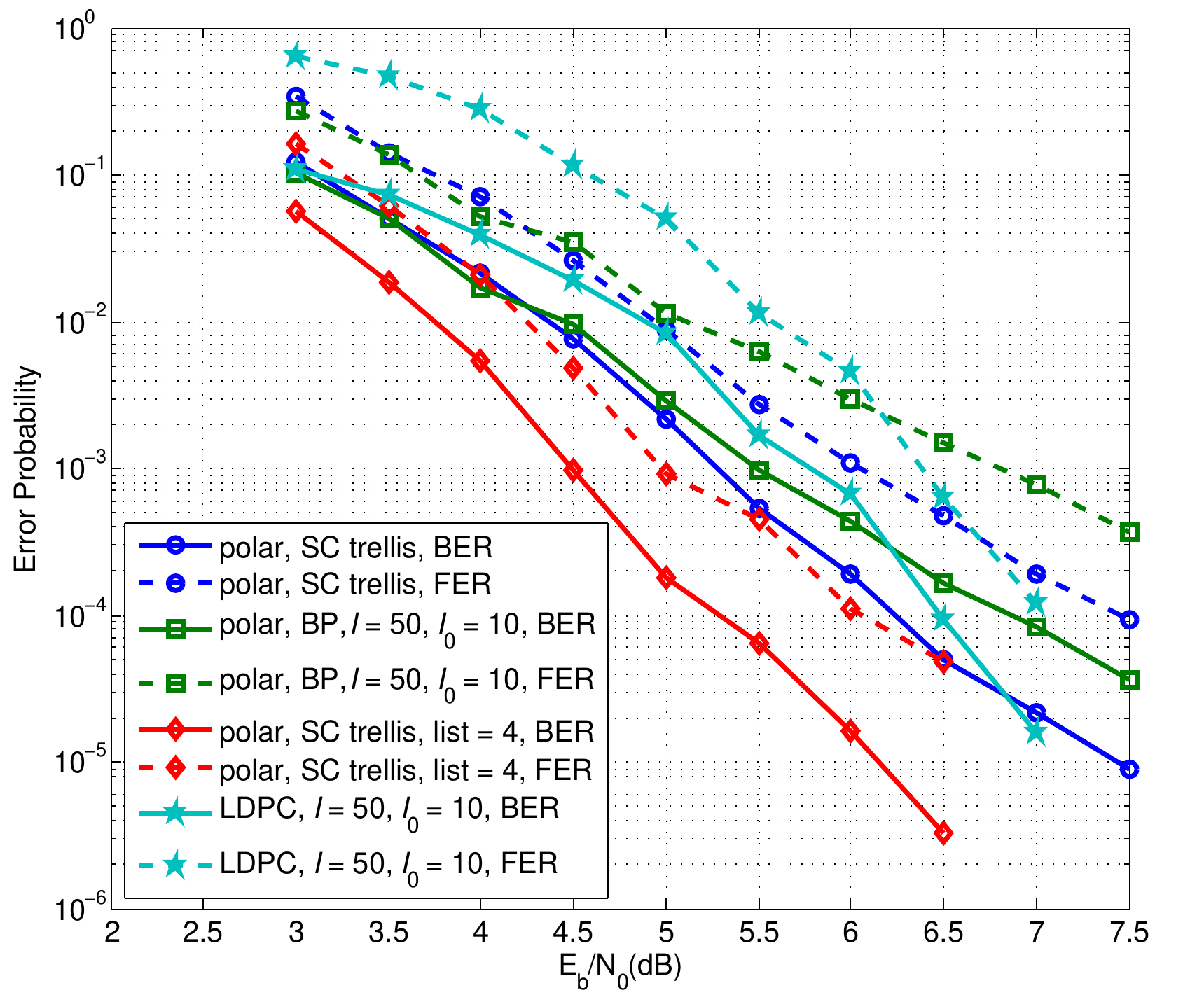}
\caption{The performance of polar codes under SC trellis decoding and BP decoding and LDPC codes under BP trellis decoding and SC decoding with code length $N=256$, code rate $R = 1/2$}
\label{Fig.: EPR4_N256_BP}
\end{figure}

We show another comparison in Fig. \ref{Fig.: EPR4_N256_BP}. 
The code lengths are all $N=256$, and code rates are $R=1/2$. In this simulation, list decoding using the proposed method is compared with the LDPC code and the proposed SC trellis decoding method is compared with the turbo equalization structure using BP decoding \cite{polar_BP}. The performance of the list decoding with 4 list paths obtain 1 dB gain at BER $10^{-4}$ as compared to the LDPC code and the SC trellis decoding method obtain more than 0.5 dB gain at BER $10^{-4}$ as compared to the turbo equalization structure using BP decoding.

Furthermore, the performance of the proposed method will be enhanced if a more suitable frozen set is selected, especially at high $E_b/N_0$ regions.

\section{Conclusion}
In this paper, we proposed SC trellis decoding to deal with the performance degradation of polar codes in ISI channels.
We show the recursive formulas of the SC trellis decoding by introducing the trellis states. Using the recursive formulas, we can decode the polar codes in ISI channels like decoding it in B-DMCs without iterations.
The error propagation can be reduced by the method. Furthermore, the proposed decoding method can be easily extended to list decoding, which gives an outstanding performance.
The simulation shows that the decoding method perform well and significantly outperforms the conventional decoding schemes.


%
%
%
%

\ifCLASSOPTIONcaptionsoff
  \newpage
\fi

\bibliographystyle{IEEEtran}
\bibliography{polar_ISI}

\end{document}